\documentstyle[aps,epsf]{revtex}
\begin{document}
\draft
\def\be{\begin{equation}}
\def\ee{\end{equation}}
\def\ba{\begin{eqnarray}}
\def\ea{\end{eqnarray}}
\title{Generalized bit cumulants for chaotic systems:\\ 
Numerical results}
\author{{Renuka Rai $^1$} \thanks{e-mail: rrai@mailcity.com}
{\it and} {Ramandeep S. Johal $^2$} 
\thanks{e-mail: raman\%phys@puniv.chd.nic.in}}
\address{{$^1$ \it Department of Chemistry, Panjab University,} \\
{\it Chandigarh -160014, India. }\\ 
 {$^2$\it Department of Physics, Panjab University,} \\
{\it Chandigarh -160014, India. }}
\date{\today}
\maketitle
\begin{abstract}
We propose generalized bit cumulants for chaotic systems,
within nonextensive thermodynamic approach. In this work, we 
apply the  first and second generalized cumulants 
to one dimensional logistic and logistic-like family of maps.
\end{abstract}
\pacs{05.45.+b, 05.20.-y, 05.70.Ce}   
\section*{I. INTRODUCTION}
Bit cumulants offer a convenient characterization of the fluctuating bit
numbers of probability distributions generated by chaotic systems 
\cite{1}. Especially, the second bit cumulant which measures 
variance of bit number, is equivalent to heat capacity in the thermodynamic
analogy. This quantity is also helpful to discuss sensitivity to
correlations among subsystems \cite{2}.

Our purpose here is to generalize the bit cumulants within nonextensive
thermostatistics of Tsallis \cite{3}. The latter formalism is based 
on a non-logarithmic entropy
\be
S_q=\frac{1-\sum_{i=1}^{W} p_i^q}{q-1},
\ee
where $\vert 1-q \vert$ is a measure of nonextensivity of the entropy
{\it i.e.} its feature of non-additivity with respect to entropies
of statistically independent subsystems. 
Based on the idea of q-deformed bit numbers \cite{3a,3pla}, Tsallis entropy
can be written in two equivalent forms
\be
S_q=-\sum_{i=1}^{W} [a_i]p_i =\sum_{i=1}^{W}[b_i]p_i^q, \label{2}
\ee
where $b_i=-a_i$ is the bit number and $[x]=\frac{q^x-1}{q-1}$.
As $q\rightarrow 1$, $[x]\rightarrow x$ and $S_q\rightarrow 
-\sum p_i{\rm ln}\;p_i$, the Shannon entropy. Thermostatistics based on this
formalism obeys the Legendre Transform structure of the
standard formalism \cite{3b}. Apart from this, various (in)equalities of thermodynamics
are properly generalized or are left invariant with respect to $q$.

Tsallis formalism has found a number of significant applications, 
such as stellar polytropes \cite{4}, two-dimensional pure
electron plasma turbulence \cite{5}, solar neutrinos \cite{6},
anomalous diffusion \cite{7}, dynamical response theory \cite{8},
to name only a few. These systems are characterized by one of the
following: long range interactions, long term memory effects
or multifractal-like phase space. Particularly, Tsallis 
formalism has yielded important insights into low-dimensional
dissipative systems at the onset of chaos or at bifurcation
points \cite{9,10}. Recently, a nonextensive thermostatistics 
based on multifractal formalism was developed by the authors which
related degree of nonextensivity $(1-q)$ to the precision of
a calculation \cite{11a,3a}. Thus there is relevance to discuss 
the alternative tool of bit cumulants within nonextensive
approach.

This paper is organized as follows: in section II we briefly discuss
the standard bit cumulants. In section III, we present generalized
version of bit cumulants and apply the first and second cumulant
to logistic-like family of maps. Section IV concludes the study.

\section*{ II.  STANDARD BIT CUMULANTS}
Bit cumulants $\Gamma_k$ of order $k$ are defined via a 
generating function
\be
G(\sigma)={\rm ln}\;(\sum_i p_i{\rm exp}(-\sigma a_i))=
\sum_{k=0}^{\infty }\left (\frac{\sigma^k}{k!}\right) \Gamma_k.
\ee
Alternatively, we can write
\be
\Gamma_k = \frac{\partial^k}{\partial \sigma^k} G(\sigma )\bigg{\vert}
_{\sigma=0}.
\ee
The zeroth cumulant is zero. The first cumulant $\Gamma_1$ is Shannon entropy
$-<a_i>$, of the distribution $\{p_i\}$. The second cumulant is variance
of the bit number {\it i.e.}, $\Gamma^2=<a_i^2>-<a_i>^2$.
From thermodynamics point of view, $\Gamma_2$ is of major importance.
An important property of $\Gamma_k$ is that it is 
additive with respect to statistically independent systems 
or subsystems. Thus for a composite system $(I+II)$
whose joint probabilities factorize as: $p_{ij}^{(I+II)}=
p_i^{(I)} .p_j^{(II)}$,we have $\Gamma_k^{(I+II)}=\Gamma_k^{(I)}+
\Gamma_k^{(II)}$. This may be taken as {\it extensive} feature 
of the standard bit cumulants.

\section*{III.   GENERALIZED BIT CUMULANTS}
We have seen (Eq. (\ref{2})) that within Tsallis thermostatistics, 
the generalized
bit-number is given by  $-[a_i]$. Thus we define the new
generating function of bit cumulants as
\be
G^{(q)}(\sigma )={\rm ln}(\sum_i p_i {\rm exp}(-\sigma[a_i]))=
\sum_{k=0}^{\infty}\left(\frac{\sigma^k}{k!}\right) \Gamma_k^{(q)}.
\ee
The generalized bit cumulant may be defined as
\be
\Gamma_k^{(q)} = \frac{\partial^k}{\partial \sigma^k} G^{(q)}(\sigma )
\bigg{\vert}_{\sigma =0}.
\ee
It is easy to see that the first cumulant is Tsallis entropy
$-<[a_i]>$. The second cumulant is the variance of the generalized 
bit number $-[a_i]$, and is given by
\be
\Gamma_2^{(q)}= <[a_i]^2> -<[a_i]>^2 \label{biv}.
\ee
Alternately, in terms of the bit number $[b_i]$ (Eq. (\ref{2})), one can write
$\Gamma_2^{(q)}= \sum [b_i]^2 p_i^{(2q-1)}-\left(\sum_i[b_i]
p_i^q\right)^2$. 
Note that it is also possible to obtain the second cumulant, by
defining a generalized free energy $\Psi_q$  \cite{11a} and using the relation
$\Gamma_2^{(q)} = - \frac{{\partial}^2 \Psi_q}{\partial \beta^2}\bigg{\vert}
_{\beta =1}$ such that 
\be
\Gamma_2^{(q)}= q\left\{\sum [b_i]^2 p_i^{(2q-1)}-\left(\sum_i[b_i]
p_i^q\right)^2\right\}.
\label{sbc}
\ee
Thus the two cumulants differ only by factor of $q$. In the following, 
we will apply Eq. (\ref{sbc}), as it is related to the 
general thermodynamic framework as established in \cite{11a}.

An important distinctive feature of the new cumulants is that 
they are non-additive (nonextensive) with respect to independent 
subsystems. In this paper, we apply the first and second 
generalized cumulants to the study of chaotic systems, such
as logistic map and logistic-like family of maps.
\subsection*{A. First cumulant}
As said above, the first generalized bit cumulant is Tsallis   
entropy $S_q$  itself. For an ergodic map, we write $S_q=
\frac{<\rho^{q-1}-1>}{1-q}$ where $\rho$ is the natural invariant
density of the map. Consider the standard logistic map
$x_{n+1}= r x_n(1-x_n)$, $x_n=[0,1]$, which is chaotic above 
$r=r_c=3.569945\dots$. As Fig. 1 shows, for $q<1$ with decrease in box size 
$\epsilon$,
Tsallis entropy shows a corresponding increase. This behaviour
is comparable to that shown in Fig. 2 by Shannon entropy $S_1=-<{\rm ln}\rho>$,
although at a given parameter value $r$, and for $q<1$, $S_q>S_1$,
at the same box size. This latter feature is already known 
for an equiprobability distribution \cite{3a}, but here it  is 
shown for a nonuniform distribution such as generated by logistic map.
Moreover, Tsallis entropy 
provides notable variation with respect to $q$ (Fig. 3). For $q<1$, as $(1-q)$
increases, $S_q$ also shows an increase. Thus results of Fig. 1 and Fig.
3 suggest an interesting possibility. Tsallis entropy evaluated at
smaller box size and small $(1-q)$ can be matched by the value of 
entropy at large box size and large $(1-q)$ value. In other 
words, for a given value of Tsallis entropy, there can be a range of 
$(1-q)$ and box size values $\epsilon$ and it is interesting to see the
relation between the two. For concreteness, we chose a fixed 
$r$ value and evaluate $S_q$ at some box size and given $q$ value. 
Then we keep $S_q$ fixed to within good approximation and 
plot the corresponding $1-q$ and $1/V = -1/{\rm ln}\;\epsilon$ values in Fig. 4.
Note that $V$ is the volume parameter in thermodynamic analogy \cite{1}. Thus
 thermodynamic limit $V\to \infty$ is equivalent to $\epsilon\to 0$.
The direct proportionality between $(1-q)$ and $1/V$ plays
important role in the nonextensive formalism for chaotic systems \cite{11a}.
\subsection*{B. Second cumulant}
In the canonical framework, second cumulant is equivalent 
to heat capacity. In the following, we make a detailed study of 
generalized second 
cumulants. In terms of probabilities ${p_i}$, we can 
write Eq. (\ref{sbc}) as 
\be
\Gamma_2^{(q)} = \frac{q}{(q-1)^2}\left\{\sum_i p_i^{2q-1}
-(\sum p_i^q)^2\right\} \label{biv2} .
\ee
$\Gamma_2^{(q)}
$ goes to second bit cumulant,  
$\Gamma_2 = <({\rm ln}\; p_i)^2> -<{\rm ln}\; p_i>^2$ 
as $q\rightarrow 1$.

To discuss non-additive property of $\Gamma_2^{(q)}$, 
consider again a composite system (I+II), for which
$p_{ij}^{I+II}=p_i^{(I)}p_j^{(II)}$. Then
\ba
\Gamma_2^{(q)}(I+II)& = &\Gamma_2^{(q)}(I)+\; \Gamma_2^{(q)}(II)\nonumber \\
&  & -2(1-q)\left\{\Gamma_2^{(q)}(I)<[a_j]>_{II}
+\Gamma_2^{(q)}(II)<[a_i]>_I\right\}\nonumber   \\
&  &    +q(1-q)^2\left\{ <[a_i]^2>_I \;<[a_j]^2>_{II} 
-<[a_i]>_I^2 \;<[a_j]>_{II}^2\right\}.\nonumber
\ea
The non-additive feature  also indicates correlations among 
subsystems I and II when $q\neq 1$.

Now for ergodic maps,  based on Eq. (\ref{biv2}) we propose 
the generalized bit variance density or 
heat capacity given by
\be
{C_2}^{(q)}=\frac{q}{(q-1)^2}( < {\rho}^{(2q-1)}>
-<{\rho}^{ q} >^2) \label{A}.
\ee 
For $q\rightarrow 1$, we have $C_2= <({\rm ln}\rho)^2>-<{\rm ln}\rho >^2$.
We make a study of Eq. (\ref{A}) for logistic-like family of maps. These are given     
by $x_{n+1} = 1-a|x_n|^z$, $z> 1$, $0<a<2$ and $-1\ge x\le 1$.
Especially for $z=2$, we have standard logistic map in its centered 
representation. Fig. 5 shows both $C_2$ and ${C_2}^{(q)}$ vs. $a$
for $z=2$. It appears there is a kind of scaling factor between
$C_2$ and ${C_2}^{(q)}$. To check this, we plot  $C_2$
vs. ${C_2}^{(q)}$ in Fig. 6 and note that most of the points 
can be fitted to a straight line.

One can ask how this relation between $C_2$ and ${C_2}^{(q)}$
depends on the nature of map. In Fig. 7, we show results for different 
$z$ values. The scaling factor between ${C_2}^{(q)}$ and $C_2$ which is 
measured by the slope of straight line fits to the graphs 
 such as Fig. 6, shows a monotonic
decrease  with increasing $z$ value (Fig. 8).  In other words, 
the deviation
of the slope from unity  decreases with increase in $z$ value,
{\it i.e.} the function ${C_2}^{(q)}$ is {\it less sensitive} to $q$
for higher values of $z$.  

Alternatively, for a given map (fixed $z$ value), one can enquire how
the above mentioned slope changes with $q$. Naturally for $q\rightarrow
1$, ${C_2}^{(q)} \rightarrow C_2$ and the slope tends
to unity. These results are shown in Fig. 9.
\section*{IV. CONCLUSION}
We have generalized the bit cumulants within nonextensive approach. 
In this paper, we have concentrated on properties of first and
second bit cumulants. We have seen how keeping Tsallis entropy 
(first cumulant) constant, we get a connection between box size 
(which represents precision of a calculation) and degree of 
nonextensivity $1-q$. Secondly, we have done detailed study on
second bit cumulant applying it to logistic-like family of maps.
We note  that for large $z$, ${C_2}^{(q)} \rightarrow C_2$. 
In the light of this, we would like to point out 
a feature seen in studies on sensitivity to initial conditions 
in similar systems \cite{10}. There as $z\to \infty$, the
nonextensivity index $q\to 1$. Further work elucidating this
connection will be welcome.
\section*{ACKNOWLEDGEMENTS}
RR would like to thank University Grants Commision, India for
grant of Senior Research Fellowship.
%\begin{thebibliography}{999}

%\end{thebibliography}
\newpage
\pagestyle{empty}
\begin{figure}
\caption{ Tsallis entropy for logistic map vs. nonlinearity parameter
$r$ of the map for three different partitionings of unit interval:
1024, 2048 and 4096, number of boxes respectively. Curve with 
symbols is for 1024 boxes, $q$ is set at 0.9. There is overall
increase in Tsallis entropy in chaotic regions, as no. of
boxes increase (box size decreases).  }
\epsfbox{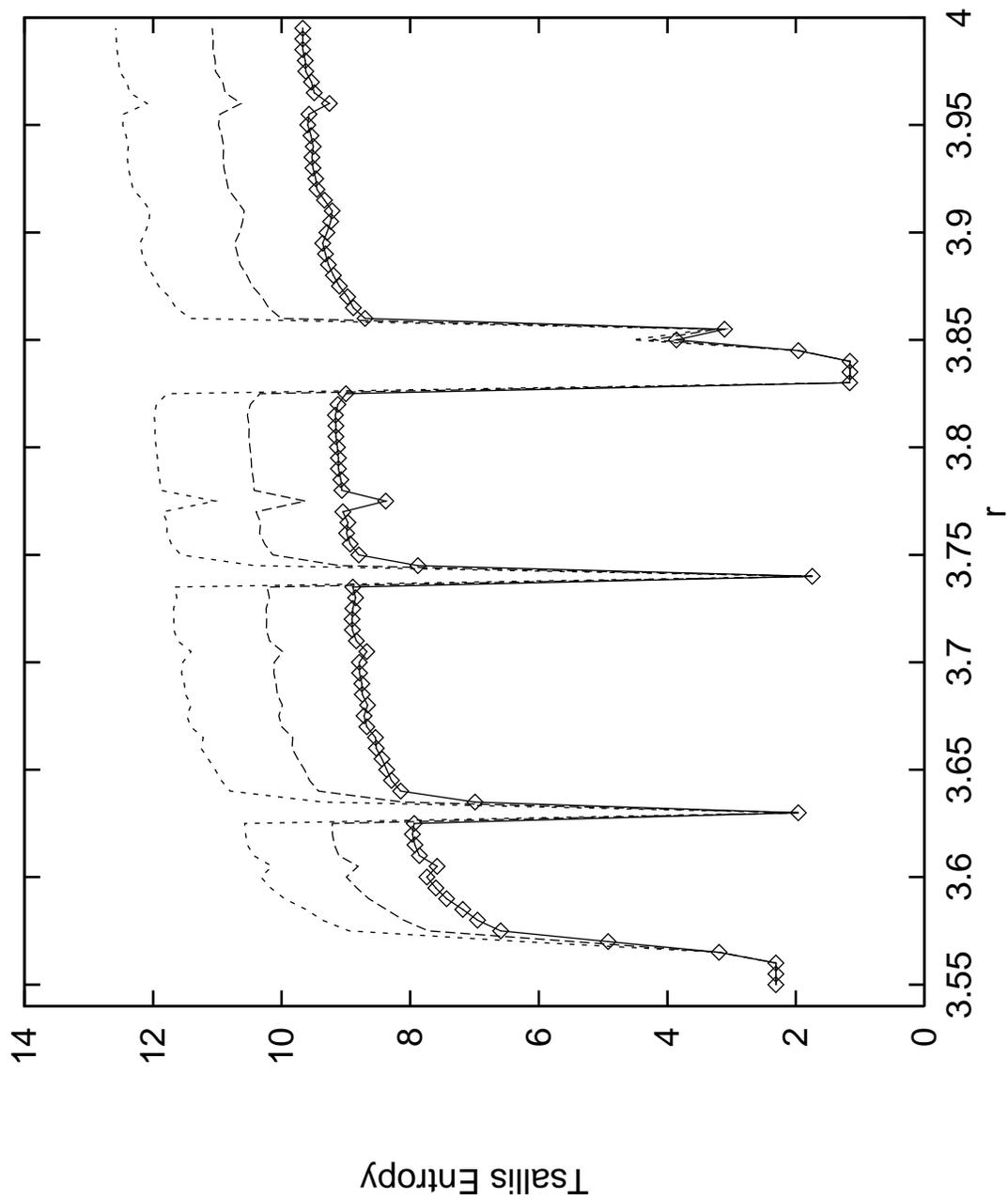}
\clearpage
\end{figure}
\begin{figure}
\caption{ Shannon entropy  vs. $r$ for logistic map and same
partitioning as in Fig. 1a. Curve with symbols is for 1024 boxes.
We see Shannon entropy also increases as box size decreases, although
its value remains lower than Tsallis entropy ($q<1$) for the same
box size.}  
\epsfbox{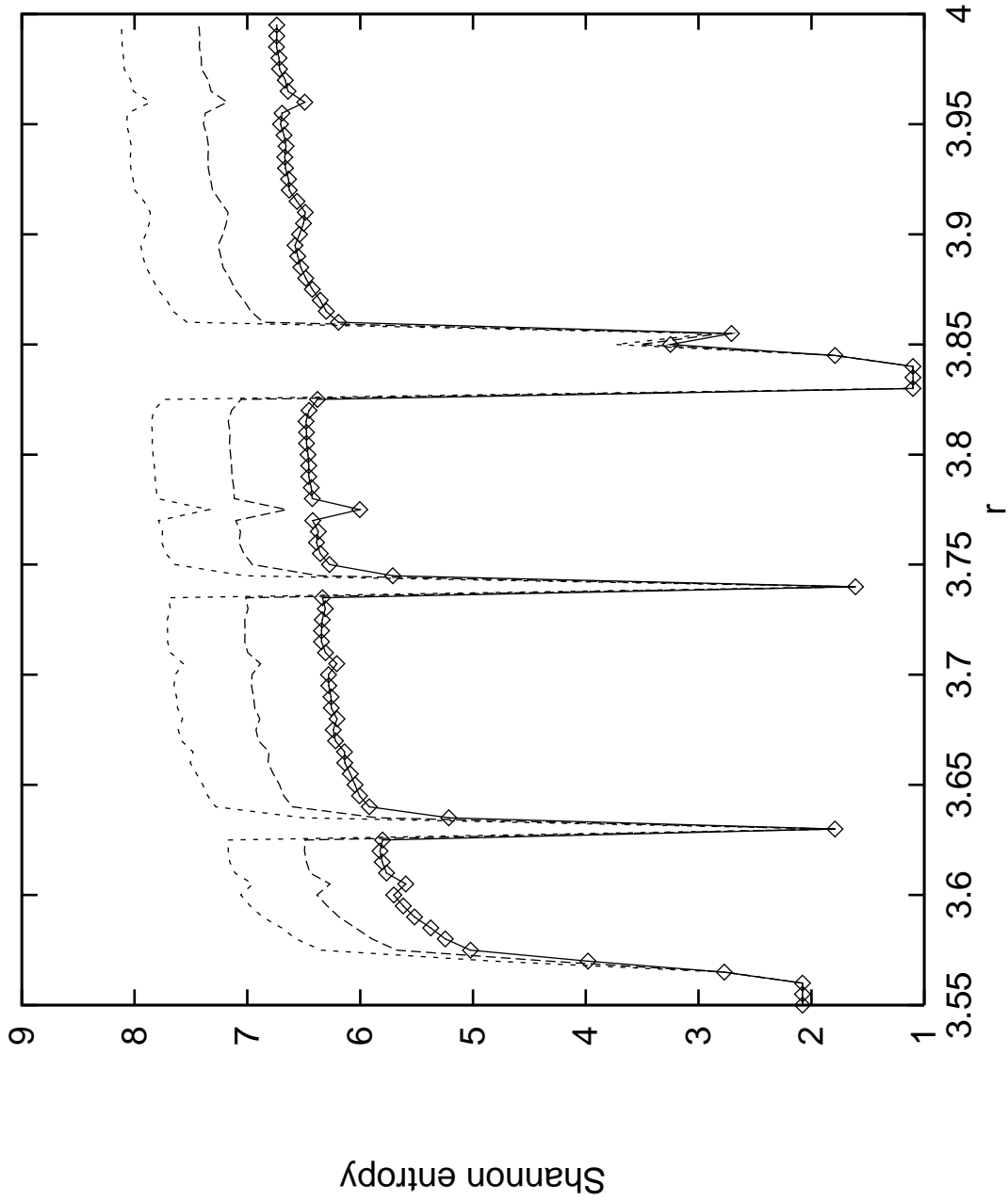}
\clearpage
\end{figure}
\begin{figure}
\caption{Tsallis entropy for logistic map vs. $r$, for different
$q$ values: 0.9, 0.93, 0.96 respectively. Number of boxes is fixed
at 1024. As $q$ increases, entropy decreases.}
\epsfbox{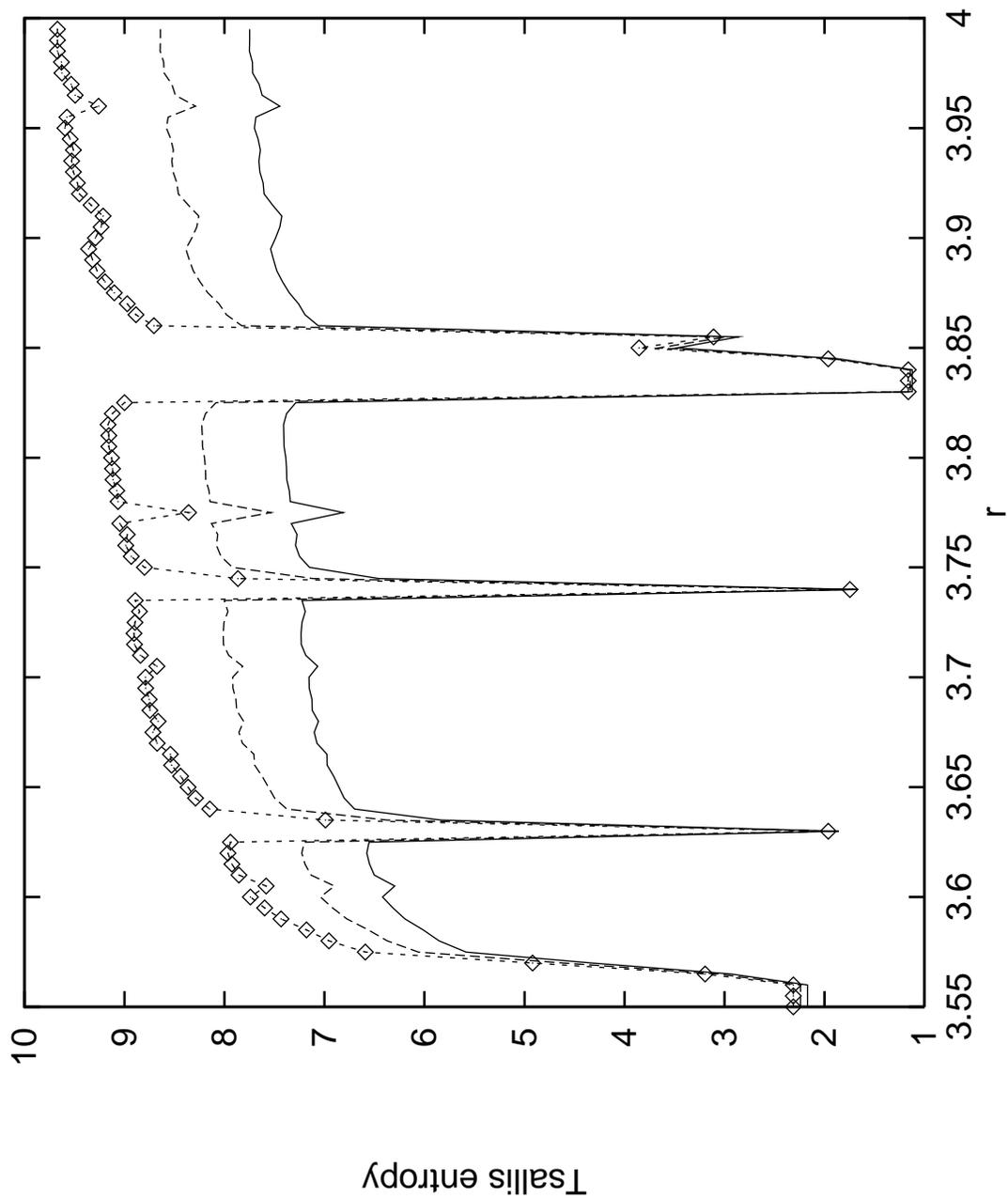}
\clearpage
\end{figure}
\begin{figure}
\caption{Relation between $(1-q)$ and $1/V$ using logistic map with 
$r =3.81$ and keeping Tsallis entropy fixed, which is taken to be 
9.15833 at $q= 0.9$. Number of boxes is 1024.}
\epsfbox{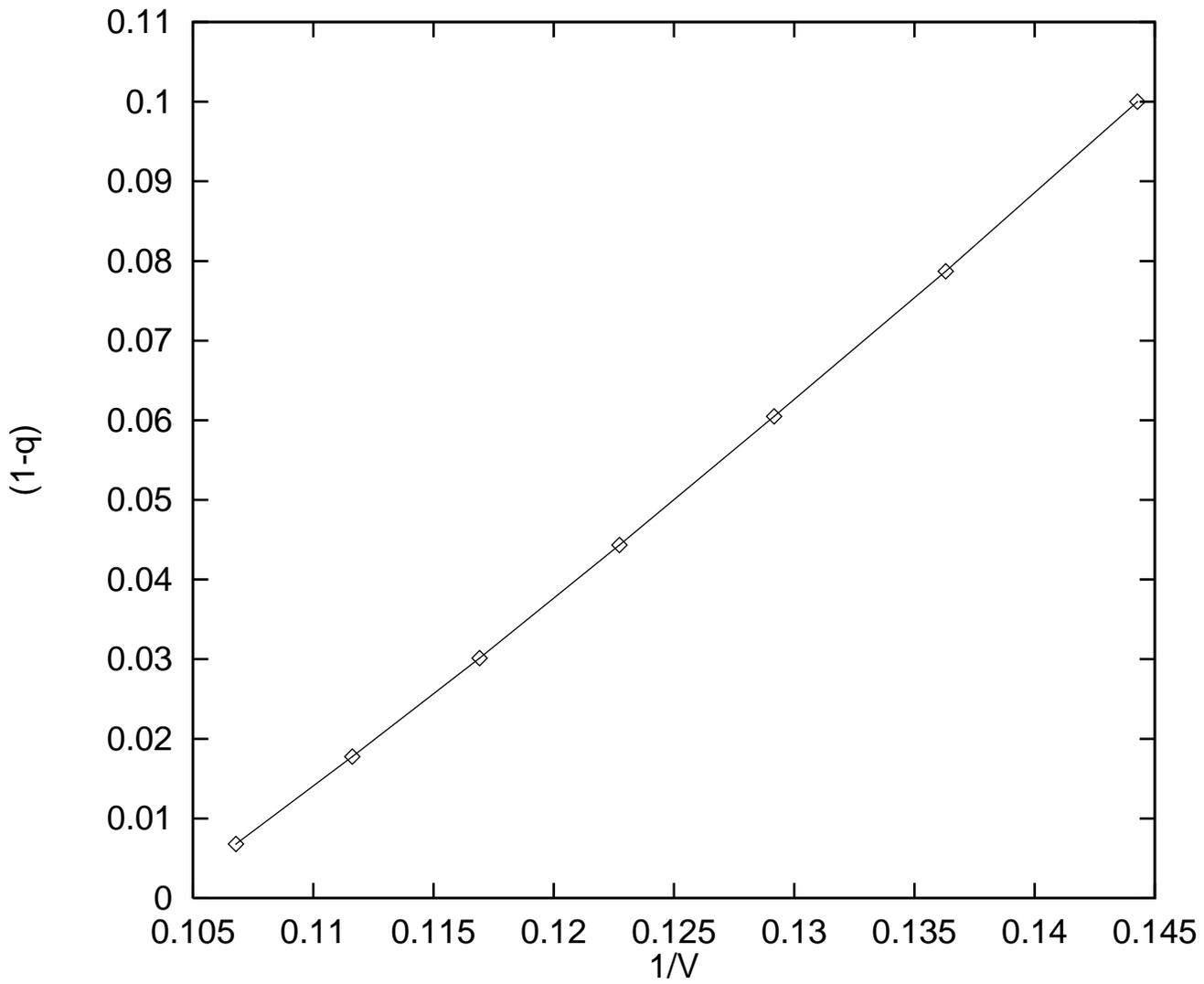}
\clearpage
\end{figure}
\begin{figure}
\caption{Heat capacity equivalent to bit variance is plotted against
control parameter $a$ of logistic map ($z=2$). Solid curve represents
$q=1$ case. Dashed curve is obtained by setting $q =0.97$ and using Eq. (\ref{A}).
Higher values of variance in nonextensive ($q\ne 1$) case can be
interpreted as due to (negative) nonextensive correlations.} 
\epsfbox{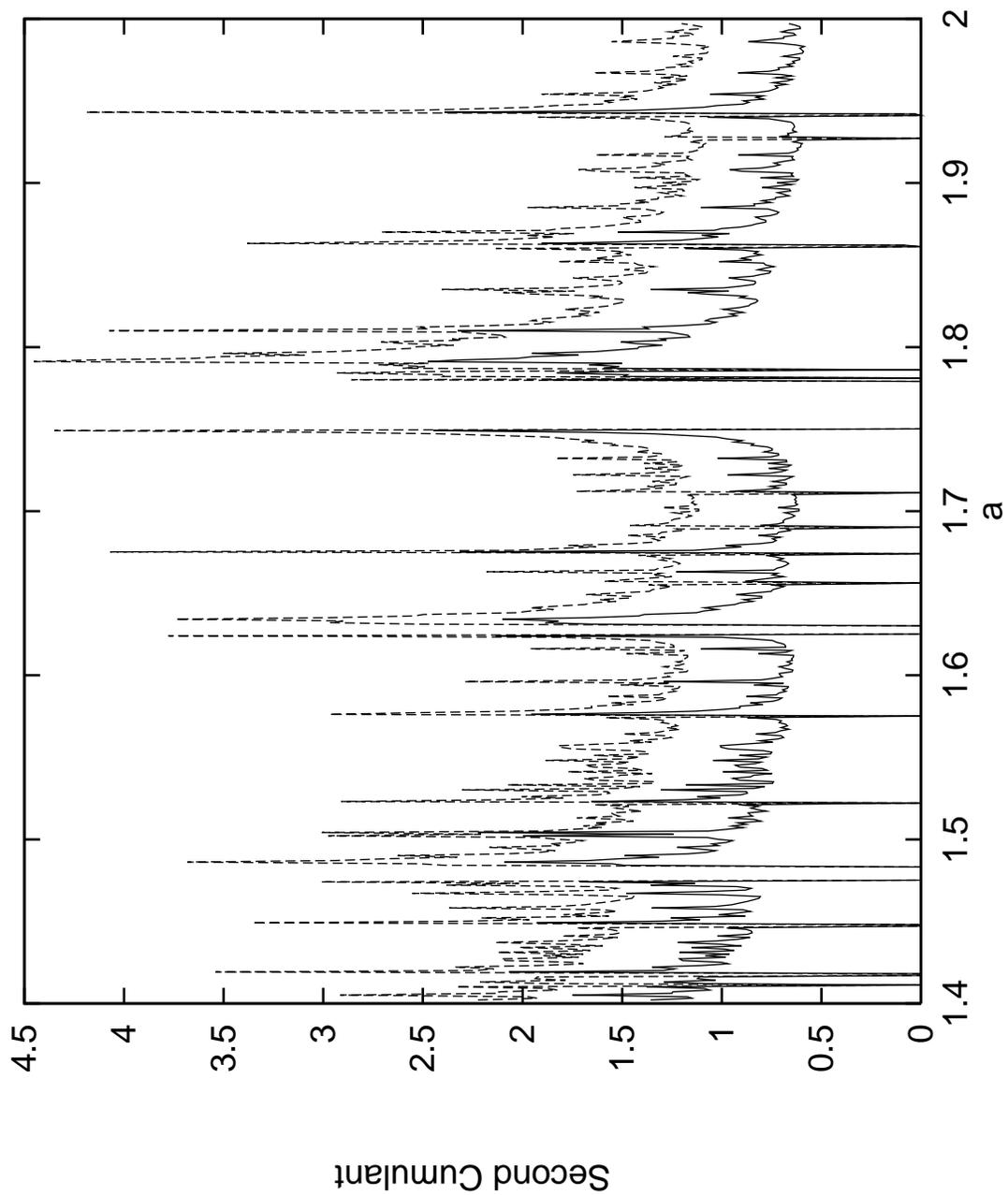}
\clearpage
\end{figure}
\begin{figure}
\caption{Scaling factor between $C_{2}^{(q)}$ and $C_2$ of Fig. 5, as
given by the slope of the straight line  fit to the data points.} 
\vskip 6pt
\epsfbox{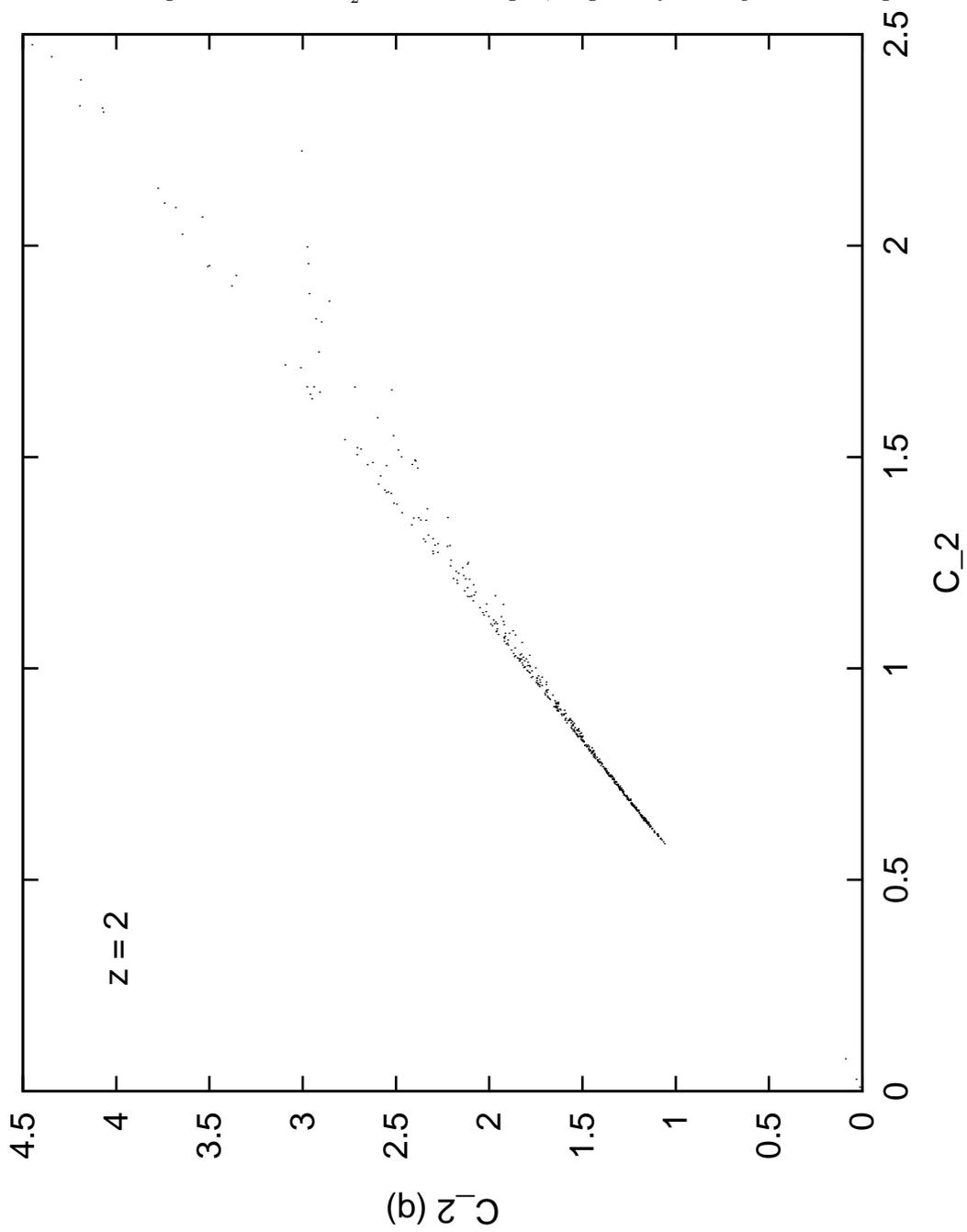}
\clearpage
\end{figure}
\begin{figure}
\caption{$C_2$ vs. $C_{2}^{(q)}$ plots for logistic-like family of maps.
(a) $z=1.5$, (b) $z=3$, (c) $z=4$, and (d) $z=5$. 
$q$ is fixed at 0.97.  Again the slope is a measure
of  scaling factor between $C_2$ and $C_{2}^{(q)}$. See also Fig. 8.}
\epsfbox{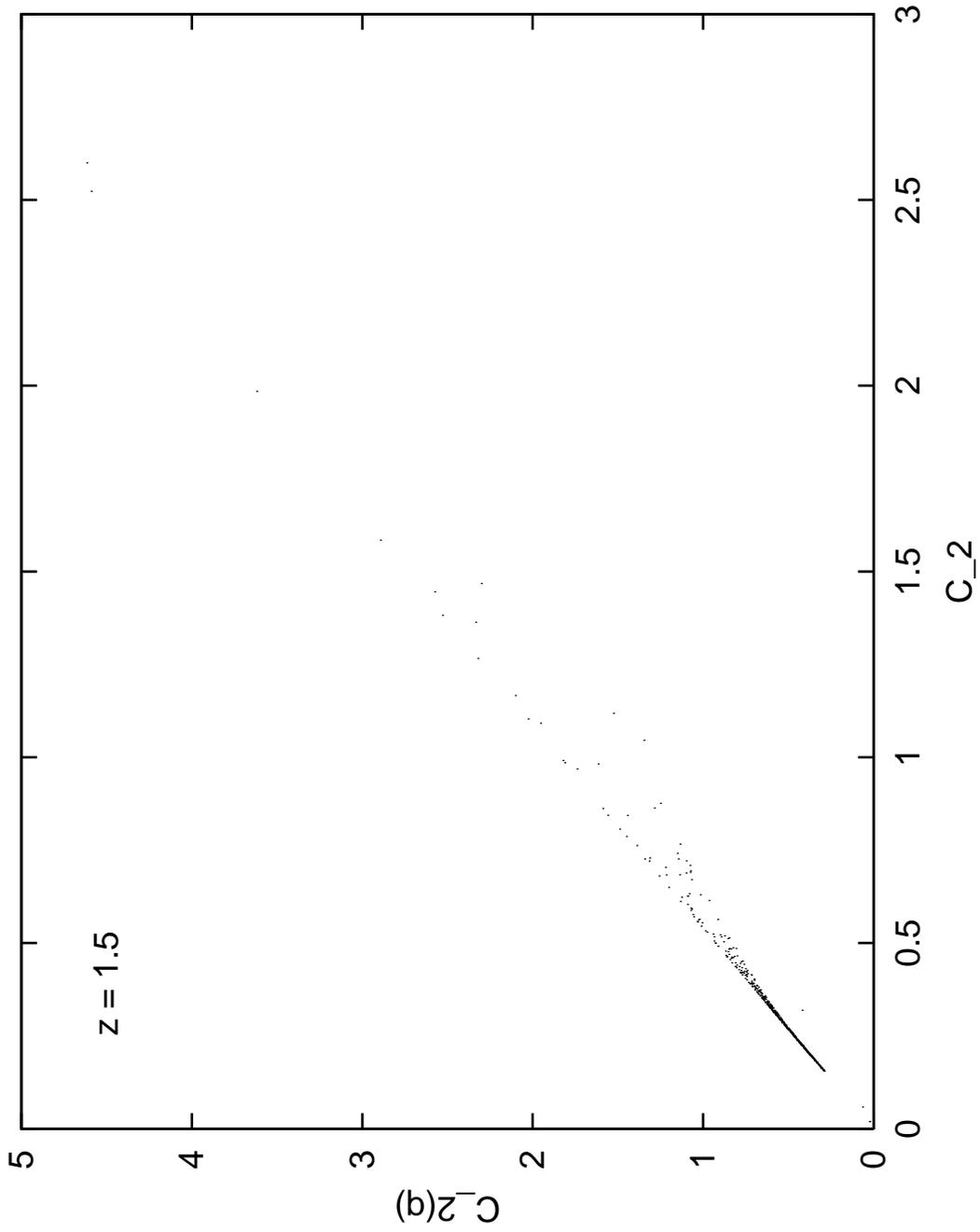}
\clearpage
\end{figure}
\begin{figure}
\epsfbox{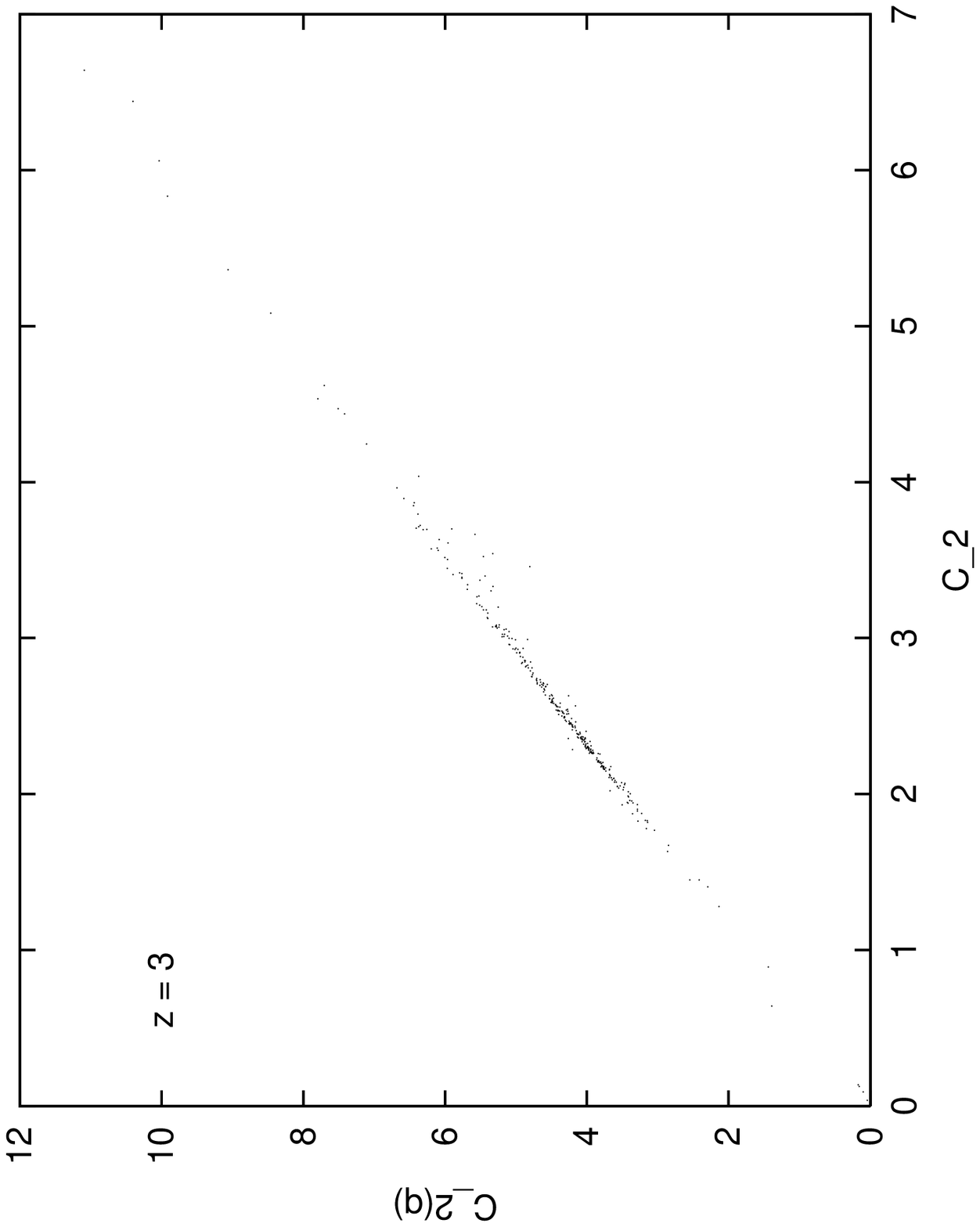}
\clearpage
\end{figure}
\begin{figure}
\epsfbox{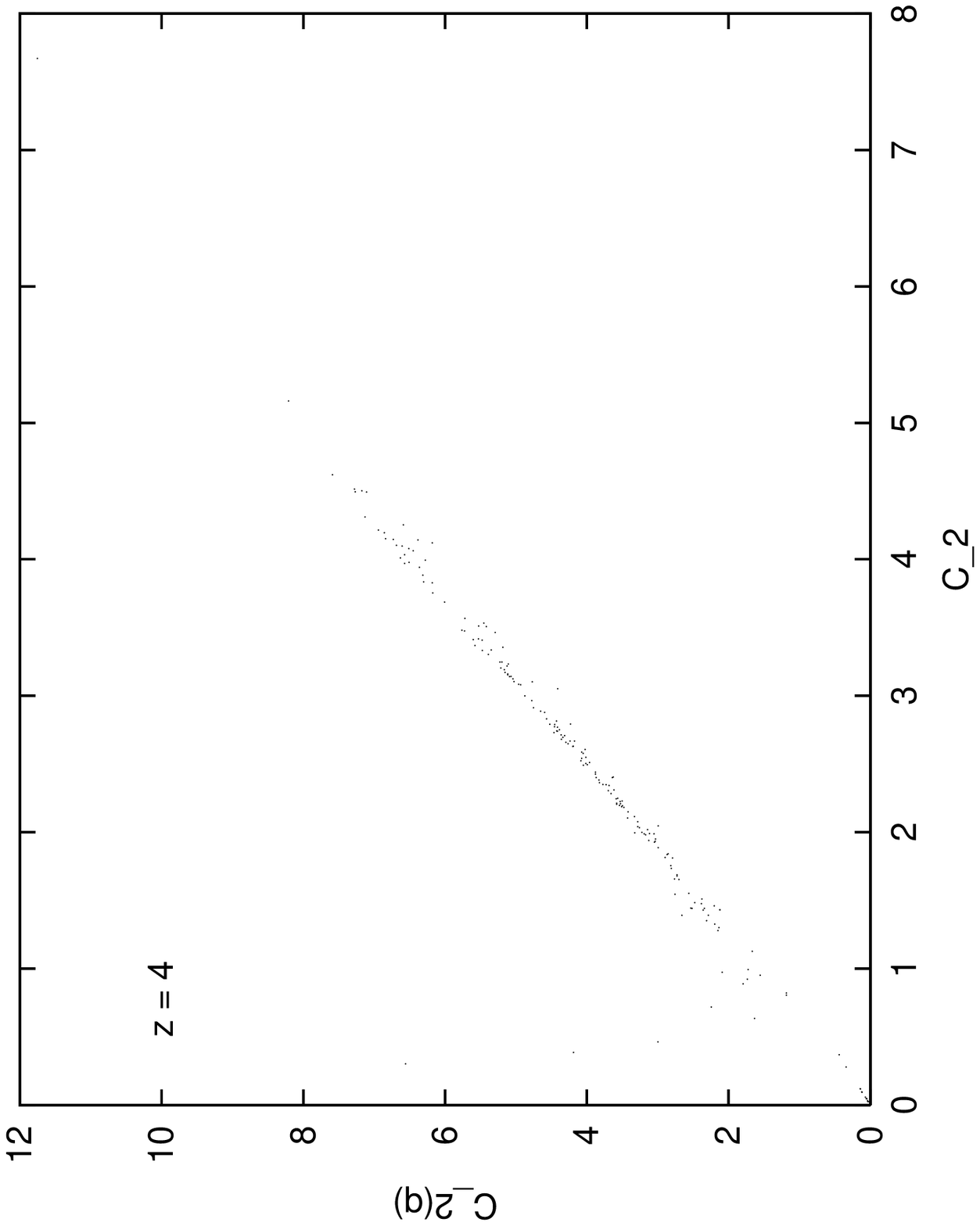}
\clearpage
\end{figure}
\begin{figure}
\epsfbox{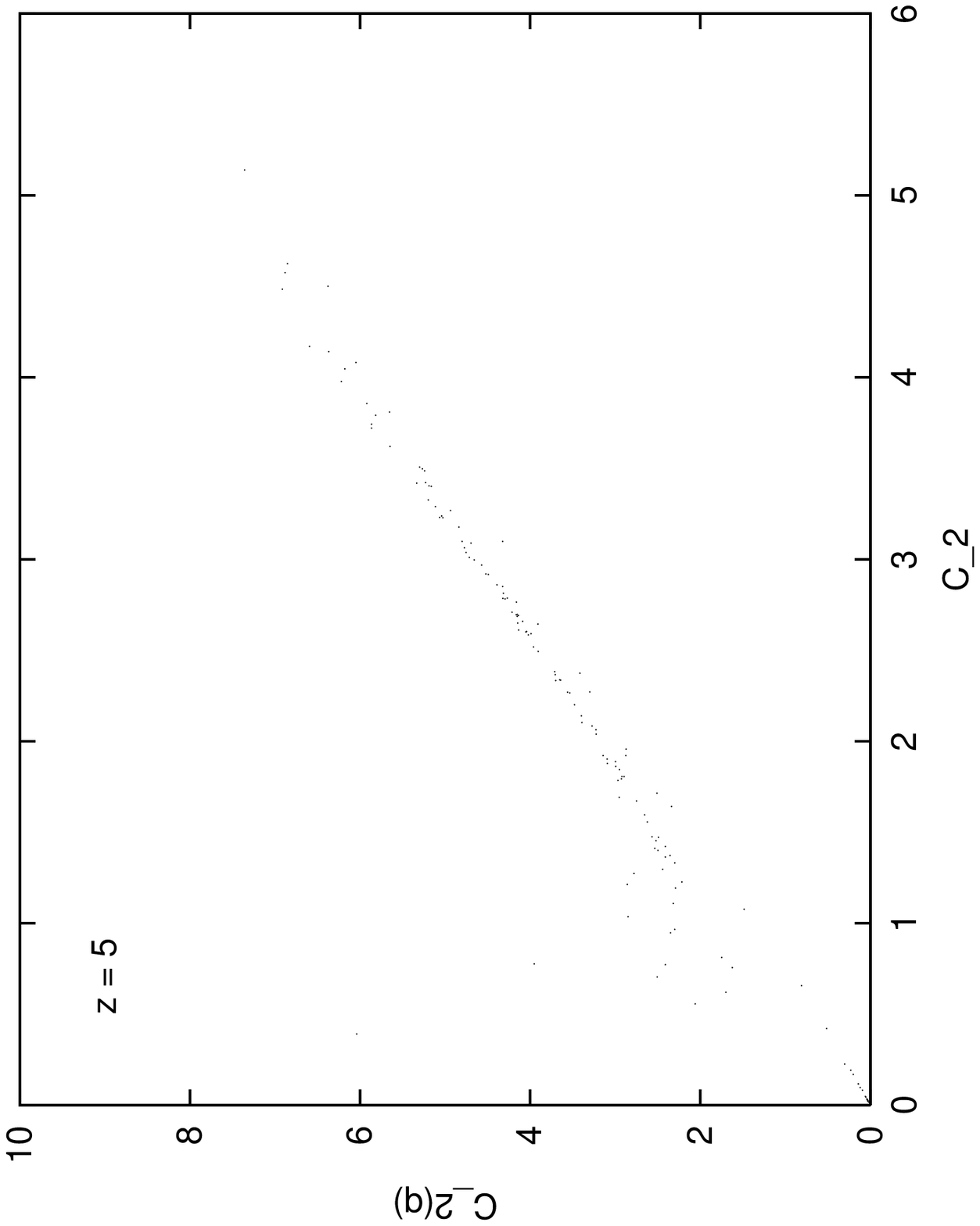}
\clearpage
\end{figure}
\begin{figure}
\caption{Behaviour of slope as obtained from Fig. 7, against nature 
of map ($z$ value).}
\epsfbox{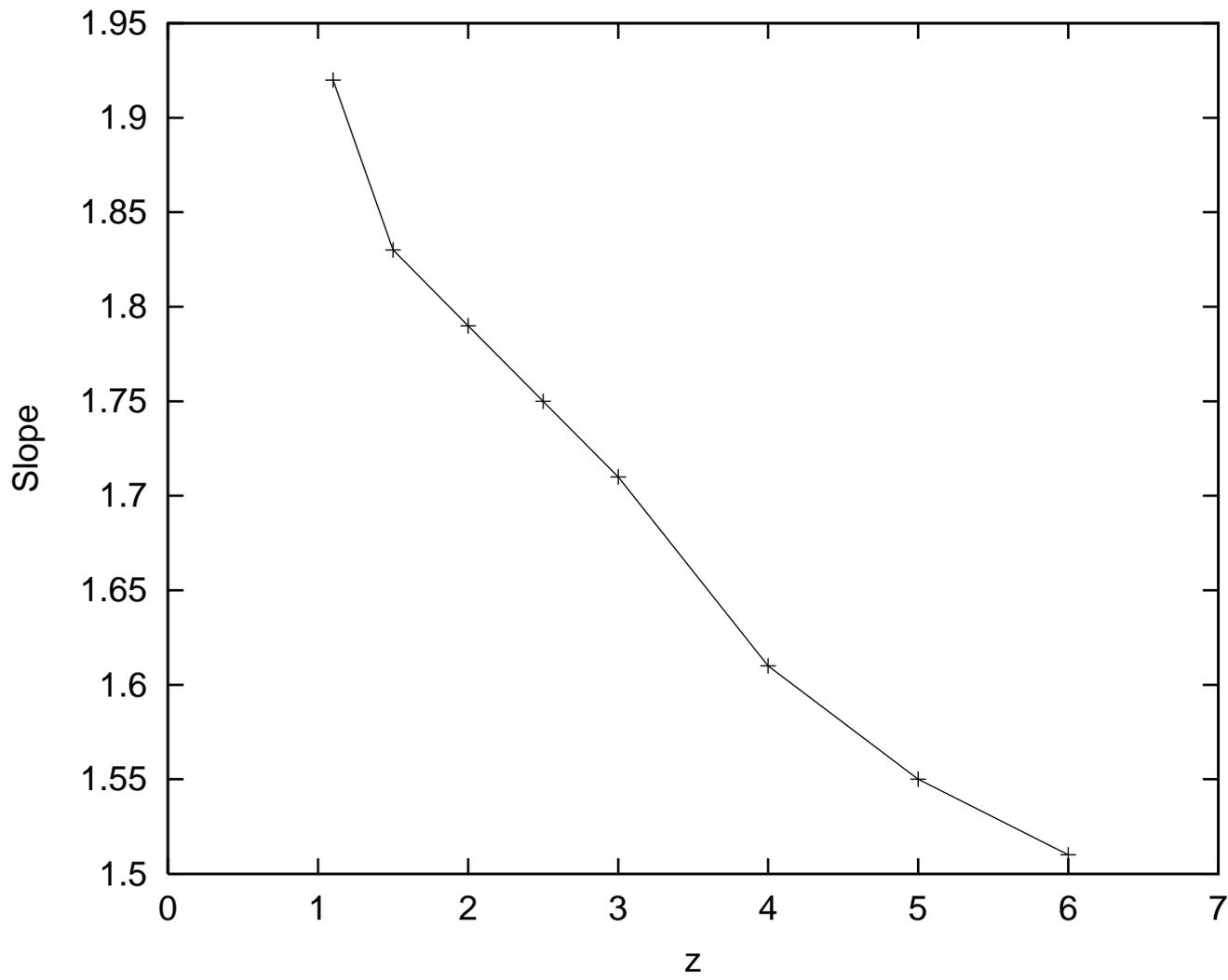}
\clearpage
\end{figure}
\begin{figure}
\caption{For a given logistic-like map (fixed $z$ value) slope vs.
$q$ value. As $q$ tends unity, slope monotonically decreases to unity.}
\epsfbox{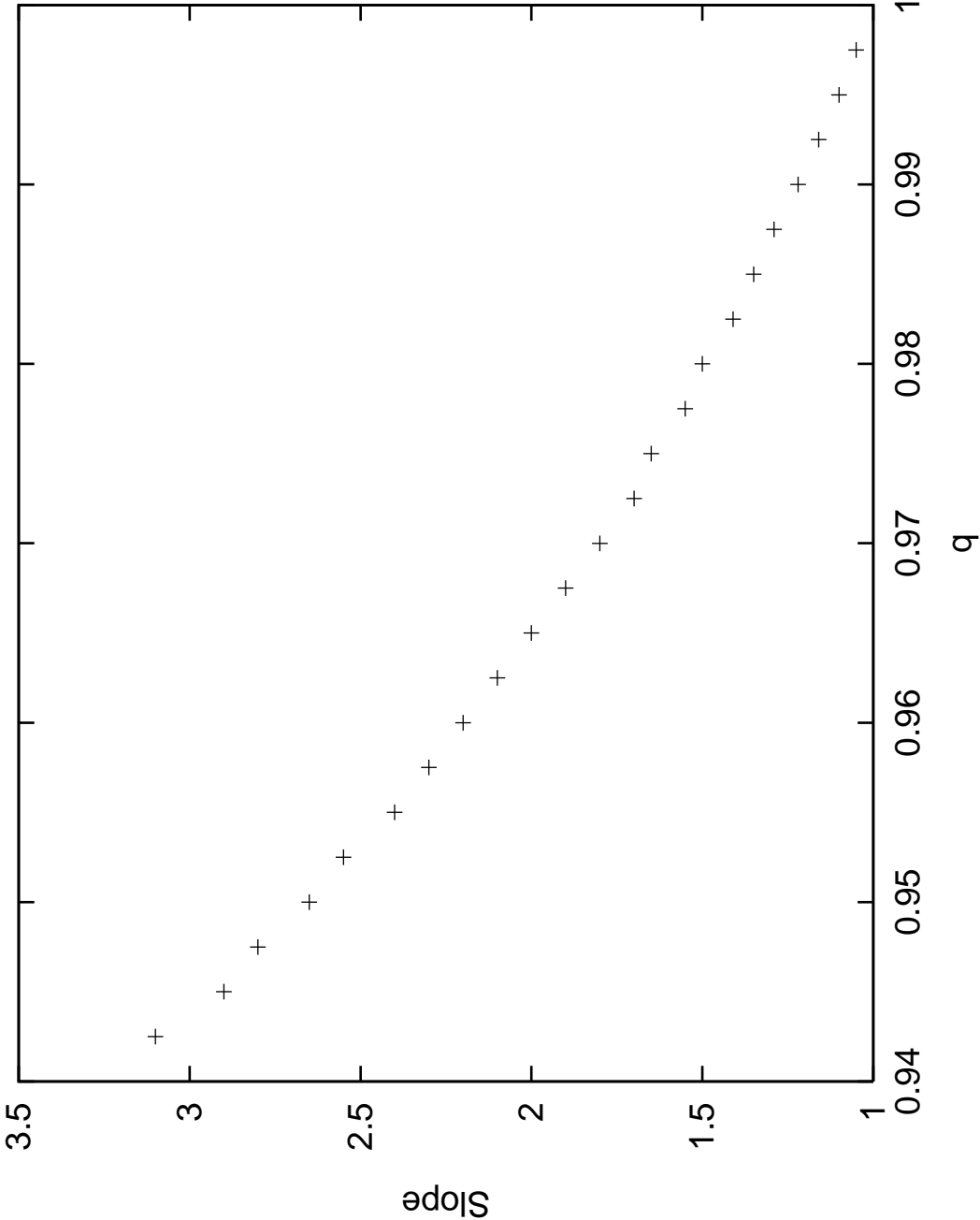}
\clearpage
\end{figure}
\end{document}